\documentclass[11pt]{article}
\usepackage{amsmath}
\usepackage{mathrsfs}
\usepackage{braket}
\usepackage{ascmac}
\usepackage{bm}
\usepackage[dvipdfmx]{graphics}
\usepackage[dviout]{graphicx}
\usepackage[at]{easylist}
\usepackage{here}
\usepackage{cite}
\usepackage{amssymb}
\usepackage{mhchem}
\usepackage{physics}
\usepackage{authblk}
\usepackage{indentfirst}
\usepackage[subrefformat=parens]{subcaption}
\usepackage[top=30truemm,bottom=30truemm,left=20truemm,right=20truemm]{geometry}
\usepackage{color}

\makeatletter
\@addtoreset{equation}{section}

\makeatother
\definecolor{DarkBlue}{rgb}{0,0,0.9} 

\definecolor{DarkRed}{rgb}{0.65,0,0}

\title{Spins of primordial black holes formed with a soft equation of state}
\date{ }
\author{Daiki Saito$^{1}$\footnote{saito.daiki.g3@s.mail.nagoya-u.ac.jp}}
\author{Tomohiro Harada$^{2}$\footnote{harada@rikkyo.ac.jp}}
\author{Yasutaka Koga$^{1}$\footnote{koga.yasutaka.k2@f.mail.nagoya-u.ac.jp}}
\author{Chul-Moon Yoo$^{1}$\footnote{yoo.chulmoon.k6@f.mail.nagoya-u.ac.jp}}

\affil{$^{1}$Division of Science, Graduate School of Science, Nagoya University, Nagoya 464-8602, Japan}
\affil{$^{2}$Department of Physics, Rikkyo University, Toshima, Tokyo 171-8501, Japan}

\begin{document}

\maketitle

\begin{abstract}

We investigate the probability distribution of the spins of primordial black holes (PBHs) formed in the universe dominated by a perfect fluid with the linear equation of state $p=w\rho$, where $p$ and $\rho$ are the pressure and energy density of the fluid, respectively.
We particularly focus on the parameter region $0<w\leq 1/3$ since the larger value of the spin is expected for the softer equation of state than that of the radiation fluid ($w=1/3$). 
The angular momentum inside the collapsing region is estimated based on the linear perturbation equation at the turn-around time which we define as the time when the linear velocity perturbation in the conformal Newtonian gauge takes the minimum value.
The probability distribution is derived based on the peak theory with the Gaussian curvature perturbation.
We find that the root mean square of the non-dimensional Kerr parameter $\sqrt{\langle a_{*}^2\rangle}$ is approximately proportional to  
$(M/M_{\textmd{H}})^{-1/3}(6w)^{-(1+2w)/(1+3w)}$, where $M$ and $M_{\textmd{H}}$ are the mass of the PBH and the horizon mass at the horizon entry, respectively.
Therefore the typical value of the spin parameter decreases with the value of $w$.
We also evaluate the mass and spin distribution $P(a_{*}, M)$, taking account of the critical phenomena.
We find that, while the spin is mostly distributed in the range of $10^{-3.9}\leq a_{*}\leq 10^{-1.8}$ for the radiation-dominated universe, the peak of the spin distribution is shifted to the larger range $10^{-3.0}\leq a_{*}\leq 10^{-0.7}$ for $w=10^{-3}$.

\end{abstract}

\tableofcontents

\section{Introduction}
\label{Intro}

In recent years, primordial black holes (PBHs) have been attracting attention as candidates for a substantial part of dark matter~\cite{Carr:2017jsz,Carr:2020gox} and possible origin of binary black holes (BHs) observed by the gravitational waves~\cite{LIGOScientific:2020stg}. 
The observational constraints on their abundance can provide us with information about inhomogeneity in the early universe.
In this paper, we focus on PBHs formed through the gravitational collapse of primordial density perturbations.
After the end of inflation, cosmological perturbations reenter the Hubble horizon and grow. 
If their amplitude is larger than a certain threshold, the perturbations become non-linear and are decoupled from the background evolution (turn-around), so that eventually collapse to PBHs. 
The threshold for PBH formation has been extensively discussed by analytical methods~\cite{Carr:1974nx,Harada:2013epa,Escriva:2019phb,Escriva:2020tak} and numerical simulations~\cite{Shibata:1999zs,Musco:2004ak,Harada:2015yda}.

In the standard situation with general relativity, an isolated black hole can be approximated by a Kerr black hole and characterized by the mass and spin parameters. 
From the gravitational wave observations, we can measure the chirp mass, the mass ratio and the effective spin $\chi_{\textmd{eff}}$ of the BH binaries. 
Therefore, in order to connect a PBH formation scenario and the observations, quantitative estimation of the masses and the spins of the PBHs would be crucial.
In this paper, we provide a procedure to evaluate the initial values of the PBH spins soon after those formation. 

The spins of the PBHs have been discussed in several works~\cite{Chiba:2017rvs,DeLuca:2019buf,Mirbabayi:2019uph,He:2019cdb,Harada:2020pzb,Chongchitnan:2021ehn,Harada:2017fjm}.
According to the results in Refs.~\cite{Chiba:2017rvs,DeLuca:2019buf,Mirbabayi:2019uph,He:2019cdb,Harada:2020pzb,Chongchitnan:2021ehn}, in the radiation-dominated universe, the effect of the tidal torque is suppressed and the expectation value of the non-dimensional spin parameter is much smaller than unity. 
On the other hand, in the matter-dominated universe, PBHs may typically have large spin parameters~\cite{Harada:2017fjm}.
The purpose of this paper is to interpolate the estimation of the PBH spin parameters in the case of the equation of state softer than the radiation fluid with the fluid approximation being valid. 
That is, in this paper, we estimate the spins of the PBHs formed in the universe filled with a perfect fluid with $0<w\leq 1/3$, where $w$ is the equation of state (EoS) parameter defined by the ratio between the pressure $p$ and energy density of the fluid $\rho$, $p=w\rho$. 
For example, PBHs could have formed during the QCD phase transition, where the EoS parameter takes $0.23\leq w\leq 1/3$~\cite{Borsanyi:2016ksw}. 
Following the same approach to Refs.~\cite{DeLuca:2019buf,Harada:2020pzb}, we evaluate the angular momentum inside the collapsing region based on the linear cosmological perturbations with the Gaussian statistics for them~\cite{Bardeen:1985tr,Heavens:1988}. 

This paper is organized as follows.
In Sec.~\ref{Ang}, we derive the expression for the angular momentum inside a collapsing region for a given profile of the fluctuations around a peak of the density perturbation. 
In Sec.~\ref{Peak}, we relate the expression of the angular momentum to the power spectrum through the typical profile of a peak based on the peak theory. 
The probability distribution function for the non-dimensional spin parameter $a_*$ is formally expressed in Sec.~\ref{Spin}, and then is estimated with specific conditions of the PBH formation and turn-around in Sec.~\ref{Assump}.  
In Sec.~\ref{Assump}, we particularly focus on the $w$-dependence of the root mean square of the spin parameter and the joint distribution of the mass and spin parameters.
Sec.~\ref{Sum} is devoted to a summary and discussion.
Throughout this paper, we use the geometrized units in which both the speed of light and Newton's gravitational constant are unity, $G=c=1$.

\section{Angular momentum in a collapsing region}
\label{Ang}

In this section, following Refs.~\cite{DeLuca:2019buf,Harada:2020pzb}, we calculate the angular momentum generated inside a collapsing region that originates from the configuration of the fluctuations around a density peak. 
The value of the angular momentum is evaluated at a certain time after the horizon entry of the fluctuation. 
We note that the angular momentum should be evaluated at the time of the turn-around, which implies the decoupling of the collapsing region from the background expansion of the universe because the total angular momentum would be conserved after that. 
Although, within the linear analysis, there is no definite criterion for the turn-around time, we will propose a plausible criterion for the turn-around time in Sec.~\ref{Assump}.
That is, we postpone specifying the turn-around time until Sec.~\ref{Assump}.

Let us write the spacetime metric in the following form:
\begin{align}
    ds^2=-\alpha^2d\eta^2+a^2\gamma_{ij}(dx^{i}+\beta^{i}d\eta)(dx^{j}+\beta^{j}d\eta), \label{eq:pert}
\end{align}
where $a=a(\eta)$ is the scale factor of the background flat FLRW spacetime
\begin{align}
    ds^2=a^2(-d\eta^2+dx^2+dy^2+dz^2). 
\end{align}
We also assume that the matter is given by a perfect fluid
\begin{align}
    T^{ab}=\rho u^{a}u^{b}+p(g^{ab}+u^{a}u^{b}), \label{eq:fluid}
\end{align}
and the EoS is $p=w\rho$ with $w$ being a constant satisfying $0<w\leq1/3$.

For a closed region $\Sigma$ on a spacelike hypersurface, we can define the angular momentum $ S_{i}(\Sigma)$ contained in the region as follows:
\begin{align}
    S_{i}(\Sigma)&:=\frac{1}{16\pi}\int_{\partial\Sigma}\epsilon_{abcd}\nabla^{c}(\phi_{i})^{d} \nonumber \\
    &=-\frac{1}{8\pi}\int_{\Sigma}R^{ab}n_{a}(\phi_{i})_{b}d\Sigma \nonumber \\
    &=-\int_{\Sigma}T^{ab}n_{a}(\phi_{i})_{b}d\Sigma.
     \label{eq:Komar}
\end{align}
Here, $n^{a}$ and $(\phi_{i})^{d}$ are the unit normal vector to the spacelike hypersurface and the rotational Killing vector, respectively. 
If there is a density peak inside $\Sigma$, $(\phi_{i})^{a}$ around the point $\vec{x}=\vec{x}_{\textmd{pk}}$ can be defined as
\begin{align}
    (\phi_{i})^{a}=\epsilon_{ijk}(x-x_{\textmd{pk}})^{j}\delta^{kl}\qty(\pdv{x^{l}})^{a}.
\end{align}

By substituting Eq.~\eqref{eq:pert} into Eq.~\eqref{eq:Komar} in a gauge with $\beta^{i}=0$, we obtain
\begin{align}
    S_{i}(\Sigma)&=-\int_{\Sigma}[(\rho+p)u^{a}u_{b}]n_{a}\epsilon_{ijk}(x-x_{\textmd{pk}})^{j}\delta^{kl}\qty(\pdv{x^{l}})^{b}d\Sigma \nonumber \\
    &\simeq (1+w)a^4\rho_{b}\epsilon_{ijk}\int_{\Sigma}(x-x_{\textmd{pk}})^{j}(v-v_{\textmd{pk}})^{k}d^3x,
     \label{eq:Komar2}
\end{align}
to the first order of the perturbation, where $v^{i}:=u^{i}/u^{0}$ with the four-velocity of the fluid $u^{a}$ and we have defined $\rho_{b}$ as the energy density in the background.

Since we aim to estimate the PBH spin, we define $\Sigma$ as a region where the matter collapses into a PBH: $\{\vec{x}|\delta(\vec{x})>f\delta_{\textmd{pk}}\}$, where $\delta_{\textmd{pk}}:=\delta(\vec{x}_{\textmd{pk}})$ and $f$ is a constant satisfying $0<f<1$.
Expanding the density perturbation $\delta$ around the peak $\vec{x}_{\textmd{pk}}$, we obtain
\begin{align}
    &\delta\simeq\delta_{\textmd{pk}}+\frac{1}{2}\zeta_{ij}(x-x_{\textmd{pk}})^i(x-x_{\textmd{pk}})^j, \\
    &\zeta_{ij}:=\left.\frac{\partial^2\delta}{\partial x^i\partial x^j}\right|_{\vec{x}=\vec{x}_{\textmd{pk}}}.
 \end{align}
Then, up to the second order of the expansion, we can regard $\Sigma$ as an ellipsoid with principal axes 
\begin{align}
    a^2_{i}=2\frac{\sigma_{0}}{\sigma_{2}}\frac{1-f}{\lambda_{i}}\nu \quad (i=1,2,3),
  \end{align}
where $\nu:=\delta_{\textmd{pk}}/\sigma_{0}$, and $\lambda_{i}$ is the eigenvalues of $-\zeta_{ij}/\sigma_{2}$ with
\begin{align}
    &\sigma^2_{j}:=\int\frac{d^3\vec{k}}{(2\pi)^3}k^{2j}|\delta_{\vec{k}}(\eta)|^2, \\
    &\delta_{\vec{k}}(\eta)=\int d^3\vec{x}\delta(\vec{x},\eta) e^{-i\vec{k}\cdot\vec{x}}.
  \end{align}

  We also expand the velocity $v^{i}$ around the peak as
 \begin{align} 
    &(v-v_{\textmd{pk}})^{i}\simeq v^{i}_{\ j}(x-x_{\textmd{pk}})^{j}, \\
    &v^{i}_{\ j}:=\left.\frac{\partial v^{i}}{\partial x^{j}}\right|_{\vec{x}=\vec{x}_{\textmd{pk}}},
  \end{align}
and obtain
  \begin{align}
    S_{i}(\Sigma)&\simeq (1+w)a^4\rho_{b}\epsilon_{ijk}v^{k}_{\ l}J^{jl}, \label{eq:Komar3}\\
    J^{jl}&:=\int_{\Sigma}(x-x_{\textmd{pk}})^{j}(x-x_{\textmd{pk}})^{k}d^3x \nonumber \\
    &=\frac{4}{15}\pi a_{1}a_{2}a_{3} \textmd{diag}(a_{1}^2,a_{2}^2,a_{3}^2).
\end{align}

In the following sections, we set $\vec{x}_{\textmd{pk}}=\vec{0}$ without loss of generality.
Then, the angular momentum can be written as
\begin{align}
    S_{i}(\Sigma)&=S_{\textmd{ref}}(\eta)s_{ei},
    \\
    S_{\textmd{ref}}&:=(1+w)a^4\rho_{b}g(\eta)(1-f)^{5/2}R_{*}^5, \\
       \vec{s}_{e}&:=\frac{16\sqrt{2}\pi}{135\sqrt{3}}\qty(\frac{\nu}{\gamma})^{5/2}\frac{1}{\sqrt{\lambda_{1}\lambda_{2}\lambda_{3}}}(\alpha_{1}\tilde{v}_{23},\alpha_{2}\tilde{v}_{13},-\alpha_{3}\tilde{v}_{12}), \\
       \alpha_{1}&:=\frac{1}{\lambda_{3}}-\frac{1}{\lambda_{2}}, \quad \alpha_{2}:=\frac{1}{\lambda_{3}}-\frac{1}{\lambda_{1}}, \quad \alpha_{3}:=\frac{1}{\lambda_{2}}-\frac{1}{\lambda_{1}}, \\
       R_{*}&:=\sqrt{3}\frac{\sigma_{1}}{\sigma_{2}}, \quad \gamma:=\frac{\sigma_{1}^2}{\sigma_{0}\sigma_{2}},
\end{align}
where
    \begin{align}
        \tilde{v}^{i}_{\,j}
        &:=-\frac{1}{\sigma_{0}}\int\frac{d^3\vec{k}}{(2\pi)^3}\frac{k^ik_{j}}{k^2}\delta_{\vec{k}}(\eta)e^{i\vec{k}\cdot\vec{x}}.
        \end{align}
$g(\eta)$ defined by
    \begin{align}
        v_{\,l}^{k}(\eta)=g(\eta)\tilde{v}_{\,l}^{k}
    \end{align}
is the time-dependent part of the velocity shear $v^{i}_{\ j}$.

\section{Typical peak profiles with the Gaussian statistics}
\label{Peak}

\subsection{General expression for an arbitrary power spectrum}

Let us consider the linear perturbations in the long wavelength limit as initial data for computation.
We assume that the physical scale of the perturbations is larger than the Hubble horizon scale at an initial time $\eta=\eta _{\textmd{init}}$, and we can apply the separate universe approach~\cite{Harada:2015yda}. 
We also assume that the curvature perturbation $\zeta_{\vec{k}}(0)$ follows a random Gaussian distribution with
\begin{align}
    \left\langle\zeta_{\vec{k}}(0)\zeta^*_{\vec{k'}}(0)\right\rangle=(2\pi)^3\delta^3\qty(\vec{k}-\vec{k'})\left|\zeta_{\vec{k}}(0)\right|^2,
\end{align}
where we have defined $\zeta$ as the curvature perturbation in the uniform density slicing.
According to Ref.~\cite{Harada:2015yda}, in the long-wavelength regime, other perturbations such as $\delta(\eta)$ and $v^{i}(\eta)$ can be written as functions of $\zeta(\eta)$. 
In particular, in the linear order, the perturbations are proportional to $\zeta(\eta)$ or its spatial derivatives.
Therefore, we can also regard $\delta$ and $v^{i}_{\ j}$ as Gaussian fields.

The density perturbation in the constant mean curvature (CMC) slicing is given as~\cite{Harada:2015yda}
\begin{align}
    \delta_{\textmd{CMC}}(\bar{x},\eta _{\textmd{init}})&=-\frac{1}{2\pi^2a^2\rho_{b}}e^{5\zeta/2}\Delta e^{-\zeta/2} \nonumber \\
 &\simeq \frac{2}{3a^2H^2_{b}}\Delta\zeta(\bar{x},0). \label{eq:delzeta}
  \end{align}
Here, we have used the Friedmann equation for the background
  \begin{align}
    \rho_{b}=\frac{3}{8\pi}H_{b}^2.
    \label{eq:BGFLRW}
   \end{align}

Then we obtain
\begin{align}
    \delta_{\vec{k},\textmd{CMC}}(\eta _{\textmd{init}})&=-\frac{2}{3a^2H^2_{b}}k^2\zeta_{\vec{k}}(0),\label{eq:delk} \\
    \therefore \hspace{5pt} \left|\delta_{\vec{k},\textmd{CMC}}\right|^2
    &=\frac{4}{9}\qty(\frac{1}{aH _{b}})^42\pi^2
     kP_{\zeta}(k),  \end{align}
where $P_{\zeta}(k)$ is the power spectrum defined by
\begin{align}
    P_{\zeta}(k):=\left|\zeta_{\vec{k}}(0)\right|^2\frac{k^3}{2\pi^2}. \label{eq:Pk}
   \end{align}
Using this definition of $P_{\zeta}(k)$ and the time dependence of the scale factor
\begin{align}
    &a_{b}\propto\eta^{\frac{2}{1+3w}}, \\
    &aH _{b}
    \simeq\dot{a}_{b}=\frac{2}{1+3w}\frac{1}{\eta}, \label{eq:aHb}
   \end{align} 
we obtain the relation 
\begin{align}
    \sigma^2_{j}
    &=\frac{4}{9}\int\frac{dk}{k}\qty(\frac{1}{aH _{b}})^4 k^{2j+4}P_{\zeta}(k)
   \end{align}
which relates the spectral moments $\sigma_{j}$ to the power spectrum.

The velocity perturbation can be expressed using the transfer function $T_{v}(k,\eta)$ as
\begin{align}
    v^{i}(\vec{x},\eta)&=-i\int\frac{d^3\vec{k}}{(2\pi)^3}\frac{k^i}{k}v_{\vec{k}}(\eta)e^{i\vec{k}\cdot\vec{x}} \nonumber \\
    &=-i\int\frac{d^3\vec{k}}{(2\pi)^3}\frac{k^i}{k}T_{v}(k,\eta)\Phi_{\vec{k}}(0)e^{i\vec{k}\cdot\vec{x}} \nonumber \\
    &=-i\frac{3+3w}{5+3w}\int\frac{d^3\vec{k}}{(2\pi)^3}\frac{k^i}{k}T_{v}(k,\eta)\zeta_{\vec{k}}(0)e^{i\vec{k}\cdot\vec{x}}, \label{eq:vi}
    \end{align} 
where we have used the relation $\Phi_{\vec{k}}(0)=-\frac{3+3w}{5+3w}\mathcal{R}_{\vec{k},\textmd{CMC}}(0)=\frac{3+3w}{5+3w}\zeta_{\vec{k}}(0)$.
See App.~\ref{Pert} for the definitions of $T_{v}$, $\Phi_{\vec{k}}$ amd $\mathcal{R}_{\vec{k}}$.
Using $\left\langle\qty(\tilde{v}_{\,l}^{k})^2\right\rangle=1$ and Eq.~\eqref{eq:vi}, we can express $g(\eta)$ as
        \begin{align}
            g^2(\eta)=\biggl<\qty(v_{\,l}^{k})^2\biggr>
            =\qty(\frac{3+3w}{5+3w})^2\int\frac{dk}{k}k^2\qty(T_{v}(k,\eta))^2P_{\zeta}(k).
            \end{align}  
    Here, we have introduced the short hand notation $\qty(v_{\,l}^{k}(\eta))^2:=v_{\,l}^{k}(\eta)v_{\,k}^{l}(\eta)$.

In order to simplify the problem, let us assume that the density perturbation has a high peak, $\nu\gg1$.
According to the peak theory~\cite{Bardeen:1985tr,Heavens:1988}, in the large $\nu$ limit, we have
\begin{align}
    &\lambda_{i}=\frac{\gamma\nu}{3}+O(1), 
\end{align}
and the principal axes of $\Sigma$ can be written as
\begin{align}
&a_{i}\simeq r_{f}:=\sqrt{6(1-f)}\frac{\sigma_{0}}{\sigma_{1}}.
\end{align}
That is to say, an overdense region with a rare peak tends to have a nearly spherical shape. 

\subsection{The case of narrow power spectrum}
\label{power}
 
Let us assume that the power spectrum can be approximated by a monochromatic power spectrum
\begin{align}
    P_{\zeta}(k)\simeq\sigma^2_{\zeta}k_{0}\delta(k-k_{0}).
\end{align}
Under this assumption, we can express $\sigma_{j}$ and $g(\eta)$ in terms of $\sigma_{\zeta}$ and $k_{0}$ as
\begin{align}
    &\sigma_{j}=\frac{2}{3}\qty(\frac{1}{aH_{b}})^2k_{0}^{j+2}\sigma_{\zeta}, \\
    &g(\eta)=\qty(\frac{3+3w}{5+3w})k_{0}|T_{v}(k_{0},\eta)|\sigma_{\zeta}.
\end{align}

According to Refs.~\cite{Bardeen:1985tr,Yoo:2018kvb}, for the narrow power spectrum with spherical symmetry, the typical form of the curvature perturbation is given by a sinc function:
\begin{align}
    &\zeta(\eta,r)=\zeta_{\textmd{pk}}(\eta)\psi(r), \\
    &\psi(r)=\frac{\sin(k_{0}r)}{k_{0}r}.
\end{align}
Since a sinc function satisfies 
\begin{align}
    \Delta\psi(r)=-k^2_{0}\psi(r)
\end{align}
and fulfills Eq.~\eqref{eq:delk}, the density perturbation can also be described by a sinc-type profile
\begin{align}
    &\delta_{\textmd{CMC}}(\eta,r)=\delta_{\textmd{pk}}(\eta)\psi(r).\label{eq:delpro} 
\end{align}
Under the truncated Taylor expansion to the second order in $y:=k_{0}r$, $\delta_{\textmd{CMC}}$ can be expressed as
\begin{align}
    \delta_{\textmd{CMC}}(\eta,r)&=\delta_{\textmd{pk}}(\eta)\qty(1-\frac{y^2}{6}),
\end{align}
and we see that the overdense region is given by
\begin{align}
    \Sigma_{O}:=\qty{\vec{x} \middle| r<r_{0}:=\frac{\sqrt{6}}{k_{0}}}.
\end{align}

We shall define the Hubble horizon entry time of the overdense region $\eta_{\textmd{H}}$ by $(aH_{b})(\eta_{\textmd{H}})r_{0}=1$ and identify $\eta_{\textmd{init}}$ as $\eta_{\textmd{H}}$. 
Then, $\sigma_{\textmd{H}}:=\sigma_{0}(\eta=\eta_{\textmd{H}})$ and $g(\eta)$ are given by
\begin{align}
    \sigma_{\textmd{H}}&=4\sigma_{\zeta},
    \label{eq:sigH} \\
    \hspace{5pt} g(\eta)&=\frac{1}{4}\qty(\frac{3+3w}{5+3w})k_{0}|T_{v}(k_{0},\eta)|\sigma_{\textmd{H}}, \label{eq:gnarrow}
 \end{align}
respectively.

\section{Spin distribution function}
\label{Spin}

\subsection{Spin distribution function and evaluation of a typical value}

The non-dimensional Kerr parameter can be written as
\begin{align}
   a_{*}&:=A_{\textmd{ref}} s_{e}, 
\end{align}
where $A_{\textmd{ref}}$ denotes the dimensionless reference spin and 
\begin{align}
   s_{e}&:=\sqrt{s_{ei}s^{i}_{e}}. 
\end{align}

At a fixed time $\eta$, $A_{\textmd{ref}}$ can be written as
\begin{align}
   A_{\textmd{ref}}(\eta)&:=\frac{S_{\textmd{ref}}(\eta)}{M^2(\eta)} \nonumber \\
   &=\frac{(1+w)a^4\rho_{\textmd{b}}g(\eta)(1-f)^{\frac{5}{2}}R^5_{*}}{M^2(\eta)},
\end{align}
where  
\begin{align}
    M(\eta)\simeq(\rho_{\textmd{b}}a^3)(\eta)\cdot\frac{4}{3}\pi r^3_{f}
\end{align}
is the mass inside $\Sigma$ at $\eta$ and is related to the mass within the Hubble horizon at $\eta=\eta_{\textmd{H}}$
\begin{align}
   M_{\textmd{H}}\simeq(\rho_{b}a^3)(\eta_{\textmd{H}})\cdot\frac{4}{3}\pi r^3_{0}
\end{align}
by the following relation:
\begin{align}
   \frac{M(\eta)}{M_{\textmd{H}}}&=\qty(\frac{a(\eta_{\textmd{H}})}{a(\eta)})^{3w}\qty(\frac{r_{f}}{r_{0}})^3 \nonumber \\
   &=(1-f)^{3/2}\qty(\frac{a(\eta_{\textmd{H}})}{a(\eta)})^{3w}, \label{eq:MR}
\end{align}
where we have used $\rho_{b}\propto a^{-3(1+w)}$. 
From this relation, we can express $A_{\textmd{ref}}(\eta)$ as 
\begin{align}
   A_{\textmd{ref}}(\eta)&\simeq(1+w)\frac{9\sqrt{3}}{M^2}a^4(\eta)\rho_{\textmd{b}}g(\eta)\qty(\frac{1}{k_{0}})^5\qty(\frac{a(\eta)}{a(\eta_{\textmd{H}})})^{5w}\qty(\frac{M(\eta)}{M_{\textmd{H}}})^{5/3} \nonumber \\
   &=\frac{9\sqrt{3}}{32\pi}\frac{(1+w)^2}{5+3w}\qty(\frac{a(\eta)}{a(\eta_{\textmd{H}})})^{1+2w}|T_{v}(k_{0},\eta)|\sigma_{\textmd{H}}\qty(\frac{ M(\eta)}{M_{\textmd{H}}})^{-1/3}.
   \label{eq:aref}
\end{align}

In the high-peak limit $\nu\gg1$, the parameter $h$ defined by
\begin{align}
   s_{e}&=\frac{2^{\frac{9}{2}}\pi}{5\gamma^6\nu}\sqrt{1-\gamma^2}h \label{eq:se}
\end{align}
obeys the universal distribution~\cite{Heavens:1988}
\begin{align}
   P_{h}(h)dh=563h^2\exp[-12h+2.5h^{\frac{3}{2}}+8-3.2(1500+h^{16})^{\frac{1}{8}}]dh. \label{eq:hfit}
\end{align}
Since we obtain $\sqrt{\langle h^2\rangle}\simeq0.419$, we can express the RMS of $s_{ei}$ as
\begin{align}
   \sqrt{\langle s_{e}^2\rangle}\simeq5.96\frac{\sqrt{1-\gamma^2}}{\gamma^6\nu}.
\end{align} 
Therefore, the RMS of the Kerr parameter is given by
\begin{align}
   \sqrt{\langle a_{*}^2\rangle}&=A_{\textmd{ref}}(\eta)\sqrt{\langle s_{e}^2\rangle} \nonumber \\
   &\simeq \frac{5.96}{8}\frac{9\sqrt{3}}{32\pi}\frac{(1+w)^2}{5+3w}\qty(\frac{a(\eta)}{a(\eta_{\textmd{H}})})^{1+2w}\sqrt{1-\gamma^2}|T_{v}(k_{0},\eta)|\sigma_{\textmd{H}}\qty(\frac{M(\eta)}{M_{\textmd{H}}})^{-1/3}\qty(\frac{\nu}{8})^{-1}, \label{eq:Sqa}
\end{align} 
where we have used $\gamma\simeq1$ for a narrow power spectrum. 
The dependence of $\sqrt{\langle a_*^2\rangle}$ on $M$ and the scale factor $a$ can be understood by the relation
\begin{align}
   a_{*i}:=\frac{S_{i}}{M^2}
   &\propto\frac{g\rho_{b}a^4r^5_{f}}{(\rho_{b}a^3r^3_{f})^2} \nonumber \\
   &\propto\frac{g}{\rho_{b}a^2r_{f}}  \nonumber \\
   &\propto \frac{k_{0}^2}{\rho_{b}a^2}\qty(\frac{M(\eta)}{M_{\textmd{H}}})^{-1/3}\qty(\frac{a(\eta)}{a(\eta_{\textmd{H}})})^{-w}  \nonumber \\
   &\propto \qty(\frac{M(\eta)}{M_{\textmd{H}}})^{-1/3}\qty(\frac{a(\eta)}{a(\eta_{\textmd{H}})})^{1+2w}.
\end{align}
Here, we have used the relation Eqs.~\eqref{eq:gnarrow} and \eqref{eq:MR} in the third line and have used Eqs.~\eqref{eq:BGFLRW} and \eqref{eq:aHb} in the fourth line.

One may be interested in the distribution of $a_*$ for given $\nu$ and $M$. 
Since the Kerr parameter can be expressed as 
    \begin{align}
            &a_*=h C(M,\nu,w), \\
            &C(M,\nu,w):=\frac{9\sqrt{6}}{80}\frac{(1+w)^2}{5+3w}\qty(\frac{a(\eta_{\textmd{H}})}{a(\eta)})^{1+2w}\sqrt{1-\gamma^2}|T_{v}(k_{0},\eta)|\sigma_{\textmd{H}}\qty(\frac{M(\eta)}{M_{\textmd{H}}})^{-1/3}\qty(\frac{\nu}{8})^{-1},  
    \end{align}
we can define the distribution of $a_*$ for given $\nu$ and $M$ as
        \begin{align}
            &P(a_{*}|M,\nu)da_{*}:=\frac{P_{h}\qty(\frac{a_{*}}{C(M,\nu,w)})}{C(M,\nu,w)N_{a}(M,\nu,w)}da_{*}, \\
            &N_{a}(M,\nu,w):=\int_{0}^{1/C(M,\nu,w)}P_{h}(h)dh.
          \end{align}
Here, we have
normalized so that the integration of $a_{*}$ over $0\leq a_{*}\leq 1$ is unity:
          \begin{align}
            \int_{0}^{1}P(a_{*}|M,\nu)da_{*}=1.
          \end{align}

\subsection{Mass-spin distribution with the critical behavior}
\label{PaM}

Let us derive a joint probability distribution of $M$ and $a_{*}$ with the critical behavior by following Ref.~\cite{Koga:2022bij}.
In the critical phenomena, when the value of $\nu$ is around the threshold $\nu_{\textmd{th}}$, $M$ obeys the scaling law
\begin{align}
    M(\nu)=KM_{\textmd{H}}\qty{\sigma_{\textmd{H}}(\nu-\nu_{\textmd{th}})}^\kappa, \label{eq:crit}
  \end{align}
where $\kappa\simeq0.75w+0.11$~\cite{PhysRevD.59.104008}.
$K$ is a coefficient that is of order unity and we set $K=1$ for simplicity. 
We solve the relation~\eqref{eq:crit} for $\nu=\nu(M)$ and obtain the probability distribution
\begin{align}
    P(a_{*},M)dadM=P(a_{*}|M,\nu)P_{\nu}(\nu)da_{*}d\nu, \label{eq:Prob}
  \end{align}
with $P_{\nu}(\nu)$ given by~\cite{Bardeen:1985tr}
\begin{align}
    P_{\nu}(\nu)\simeq\sqrt{\frac{2}{\pi}}\frac{e^{-\nu^2/2}}{\textmd{erfc}\qty(\frac{\nu_{\textmd{th}}}{2})}.
  \end{align}

\section{Explicit evaluation of the spin distribution}
\label{Assump}

\subsection{Specific conditions for PBH formation and turn-around time}

In order to explicitly evaluate the spin distribution by using the expressions derived in the previous section, we need to specify the conditions of PBH formation and the time of the turn-around.  

\begin{figure}[htbp]
    \begin{center}
     \includegraphics[clip,width=8cm]{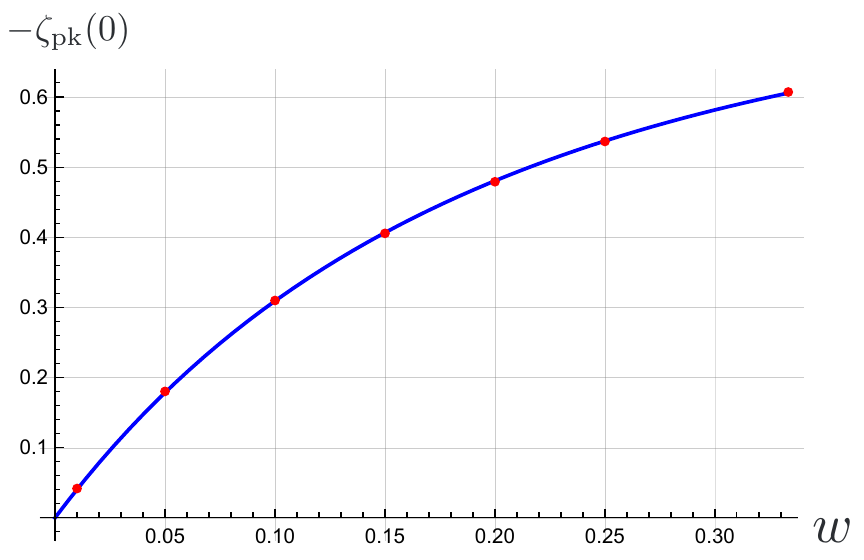}
     \caption{$w$-dependence of the thresholod value of $\zeta_{\textmd{pk}}(0)$ for a sinc-type curvature perturvation. The red points show the result of the numerical computation~\cite{Albert} and the blue line represents the fitting function~\eqref{eq:THC} with $(p_1,p_2,p_3)\simeq(1.02, 2.02, 0.991)$.}
     \label{fig:zetapeakfit}
  \end{center}
   \end{figure}

First, let us specify the PBH formation condition. 
For the fluid to collapse into a PBH, the amplitude of the perturbation must satisfy the PBH formation condition.
In this paper, we shall fix the amplitude as a threshold value for the formation.
The threshold value for a sinc-type profile obtained by the numerical computation~\cite{Albert} is shown in Fig.~\ref{fig:zetapeakfit} (red points). 
In order to obtain a simple analytic expression for the threshold value as a function of $w$, we introduce the following form of the fitting function:
\begin{align}
    \zeta_{\textmd{pk}}(0)=-p_1\sin^2\qty(\frac{p_2\sqrt{w}}{1+p_3w}). \label{eq:THC}
 \end{align}
This functional form is motivated by the analytic formula obtained by using the three-zone model~\cite{Harada:2013epa}. 
Then, we obtain $(p_1,p_2,p_3)\simeq(1.02, 2.02, 0.991)$ (blue curve in Fig.~\ref{fig:zetapeakfit}), which 
agrees with the numerical results as is explicitly shown in Fig.~\ref{fig:zetapeakfit}. 
In the following, we shall use the fitting function for the threshold value.
Then, the amplitude of the density perturbation $\delta_{\textmd{pk}}(\eta_{\textmd{H}})$ is also fixed by the relation
\begin{align}
    \delta_{\textmd{pk}}(\eta_{\textmd{H}})=-4\zeta_{\textmd{pk}}(0). 
    \label{eq:dpkzpk}
 \end{align}

Next, let us propose a plausible criterion for the turn-around time. 
In the previous section, we have not specified the time when the angular momentum should be evaluated. 
The time should be the time when the collapsing region is decoupled from
the background expanding universe. 
However, there is no definite criterion for this decoupling time within the level of the linear analysis. 
Nevertheless, we here propose a plausible criterion in order to explicitly evaluate the typical value of PBH spin.

\begin{figure}[htbp]
    \begin{center}
     \includegraphics[clip,width=10cm]{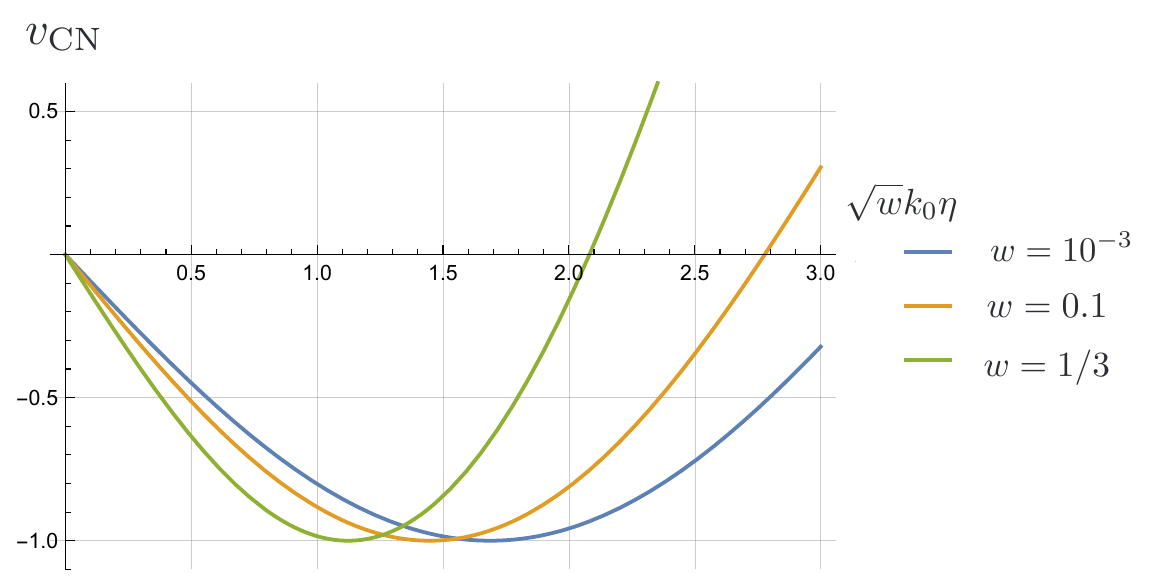}
     \caption{Time dependence of the linear velocity perturbation in the conformal Newtonian gauge $v_{\textmd{CN}}$~\eqref{eq:vCN} for $w=10^{-3}, 0.1, 1/3$.
     In these cases, we can determine the turn-around times as $k_{0}\eta_{\textmd{ta}}=16.7$, $4.57$ and $1.95$, respectively.
     In the plot, the amplitude of the perturbation is fixed so that $v_{\textmd{CN}}(\eta_{\textmd{ta}})=-1$.}
     \label{fig:vCNw}
  \end{center}
   \end{figure}

In the high-density region, after the horizon entry, the fluid element would be attracted by the gravitational potential, and accelerated inward relative to the background expansion. 
In the case in which the collapse does not halt and eventually a PBH forms, we naively expect that the gravitational force always wins against the pressure gradient force. 
However, in the linear analysis, the time derivative of the velocity perturbation takes the minimum at a certain time, and the acceleration becomes positive after this moment. 
The fact that the dynamical behavior of the linear perturbation from that expected in PBH formation may indicate the breakdown of the linear approximation and, hence, the nonlinearity would have been so significant by this moment. This implies
that the local density perturbation becomes so large that the expansion should be about to turn around. 
Therefore we here identify the following condition as the turn-around time $\eta_{\textmd{ta}}$:
\begin{align}
v'_{\textmd{CN}}(\eta_{\textmd{ta}})=0, \label{eq:TACond}
 \end{align}
where $v_{\textmd{CN}}$ is the fluid velocity in the conformal Newtonian (CN) gauge~\eqref{eq:vCN}.
In Fig.~\ref{fig:vCNw}, time dependence of $v_{\textmd{CN}}$ for $w=10^{-3}, 0.1, 1/3$ is depicited.
In these examples, the turn-around times can be computed as $k_{0}\eta_{\textmd{ta}}=16.7$, $4.57$ and $1.95$.
Since $v_{\textmd{CN}}$ obeys the linearized Euler equation~\eqref{eq:Euler}, this condition can be interpreted as no acceleration of the perturbation of the fluid. 
For the explicit evaluation in this section, we use the CN gauge, where the transfer function is given by
 \begin{align}
    T_{v_{\textmd{CN}}}(k_{0},\eta)=\frac{v_{\textmd{CN}}(\eta)}{\Phi_{k_{0}}(0)}.
 \end{align}

\subsection{Results}
\label{Res}

\subsubsection{The root mean square of the spin}
\label{Resrms}

\begin{figure}[htbp]
    \begin{center}
     \includegraphics[clip,width=8cm]{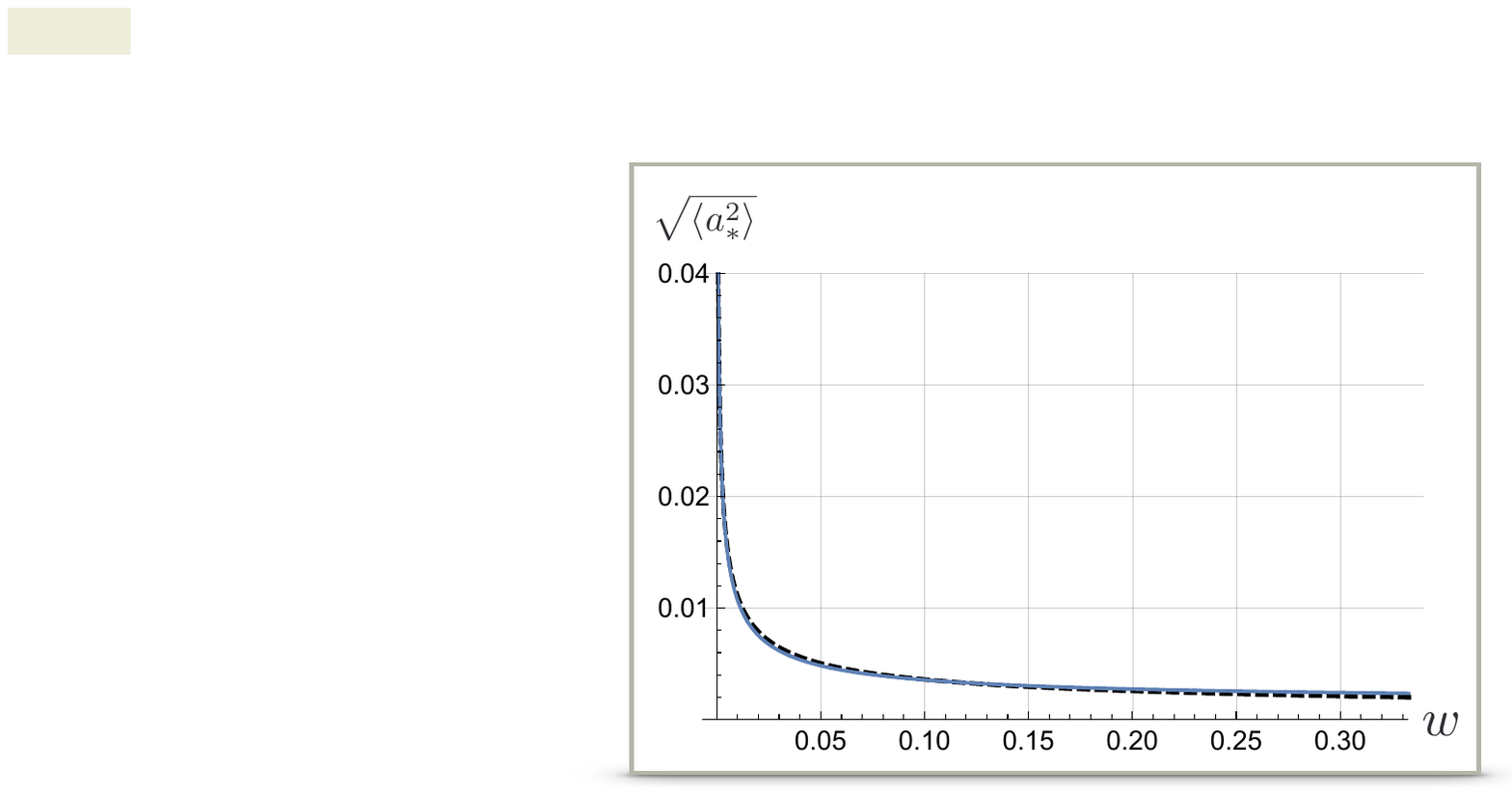}
     \caption{The RMS of the Kerr parameter $\sqrt{\langle a_{*}^2\rangle}$ as a function of $w$ (blue solid line).
     We have fixed the parameters as $\nu=8$, $M= M_{H}$ and $\gamma=0.85$. 
     The black dashed line shows the fitting function which obeys a power-law $\sqrt{\langle a_{*}^2\rangle}\propto w^{-0.49}$. }
     \label{fig:rms}
  \end{center}
   \end{figure}

Let us see the $w$-dependence of $\sqrt{\langle a_{*}^2\rangle}$. 
In order to evaluate $\sqrt{\langle a_{*}^2\rangle}$, we need to fix the parameters $\nu$, $\sigma_{\textmd{H}}$, $M$ and $\gamma$.
In the numerical analysis, we shall fix $\nu=8$, which is roughly consistent with an observationally interesting value of PBH abundance \cite{DeLuca:2019buf} and with the high-peak assumption $\nu\gg1$.
Then, the value of $\sigma_{\textmd{H}}=4\sigma_{\zeta}$ is determined automatically by the relations $\nu=\delta_{\textmd{pk}}/\sigma_{0}$, \eqref{eq:THC} and \eqref{eq:dpkzpk}, depending on $w$.
Since $\gamma\simeq 1$ for a narrow power spectrum, we fix $\gamma=0.85$.
The plot of $\sqrt{\langle a_{*}^2\rangle}$ as a function of $w$ for $\nu=8$, $M= M_{H}$ and $\gamma=0.85$ is depicted in Fig.~\ref{fig:rms}.
From the figure, we see that the RMS decreases with $w$. 
While $\sqrt{\langle a_{*}^2\rangle}\simeq2.3\times10^{-3}$ in the radiation-dominated universe ($w=1/3$), the spin increases as $w$ decreases and $\sqrt{\langle a_{*}^2\rangle}\geq O(10^{-2})$ for $w<10^{-2}$.
We also see the RMS is fitted by the power-law $\sqrt{\langle a_{*}^2\rangle}\propto w^{-0.49}$ (black dashed line). 

\begin{figure}[htbp]
    \begin{center}
     \includegraphics[clip,width=10cm]{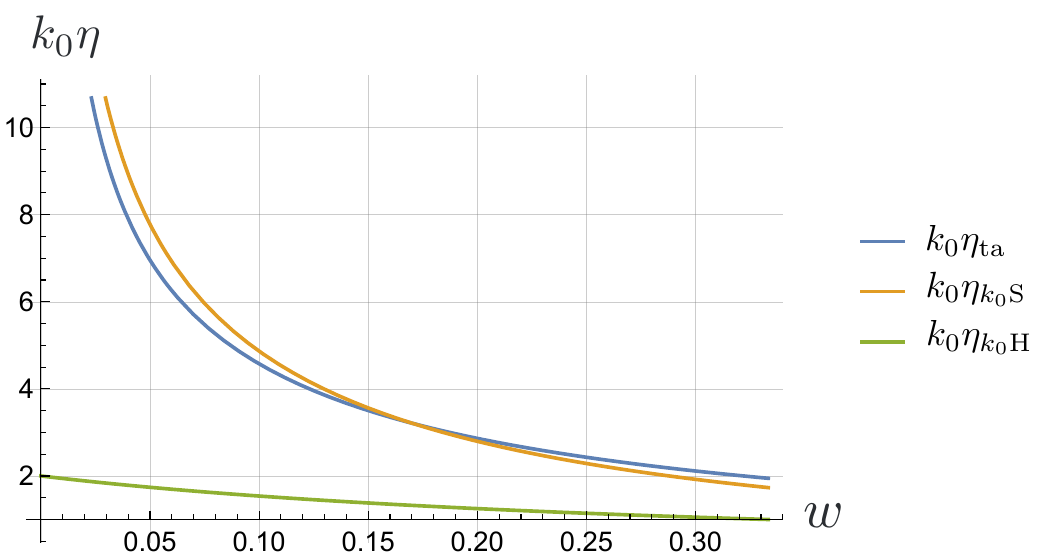}
     \caption{$w$-dependence of the turn-around time $k_{0}\eta_{\textmd{ta}}$, which is defined by the condition $v'_{\textmd{CN}}(\eta_{\textmd{ta}})=0$.
     We also plot $\eta_{k_{0}\textmd{S}}$  and $\eta_{k_{0}\textmd{H}}$, defined by the Hubble and sonic horizon entry of the inverse wave number $1/k_{0}$~\eqref{eq:khorizonentry}.}
     \label{fig:horizonk0}
  \end{center}
   \end{figure}

\begin{figure}[htbp]
    \begin{center}
      \includegraphics[width=10cm]{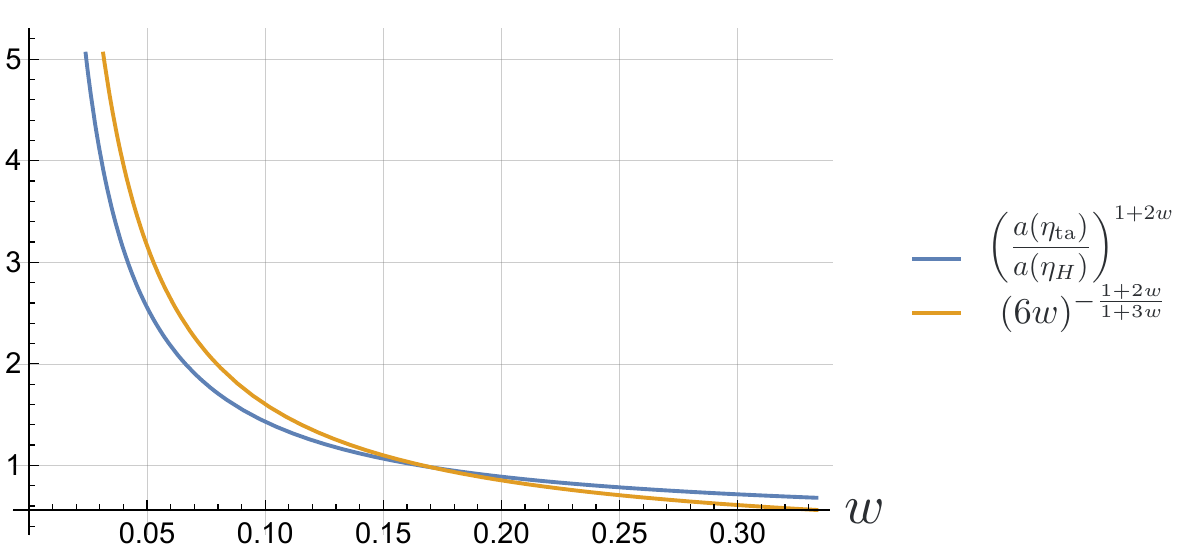}
      \caption{The blue line shows $w$-dependence of the factor in the RMS $\qty(\frac{a(\eta_{\textmd{ta}})}{a(\eta_{H})})^{1+2w}$.
      The orange line shows the ratio between the conformal time 
      $\qty(\frac{\eta_{k_{0}\textmd{S}}}{\sqrt{6}\eta_{k_{0}\textmd{H}}})^{\frac{2}{1+3w}}=\qty(6w)^{-\frac{1+2w}{1+3w}}$ 
      , which is obtained by the relation~\eqref{eq:aratio}.}
      \label{fig:aratio}
    \end{center}
  \end{figure}

We can see that the $w$-dependence of $\sqrt{\langle a_{*}^2\rangle}$ reflects that of the turn-around time $\eta_{\textmd{ta}}$.
The plot of $k_{0}\eta_{\textmd{ta}}$ is displayed in Fig.~\ref{fig:horizonk0}.
We also plot $\eta_{k_{0}\textmd{H}}$ and $\eta_{k_{0}\textmd{S}}$, defined by the Hubble and sonic horizon entry of the inverse wave number $1/k_{0}$, respectively:
\begin{align}
    \frac{aH(\eta_{k_{0}\textmd{H}})}{k_{0}}=1, \quad \frac{aH(\eta_{k_{0}\textmd{S}})}{k_{0}}=\sqrt{w},
    \label{eq:khorizonentry}
\end{align}
in the figure.
From this figure, we see that the turn-around time $\eta_{\textmd{ta}}$ has a similar $w$-dependence to $\eta_{k_{0}\textmd{S}}$.
That is to say, under the assumptions we have made, the perturbation decouples from the expansion almost simultaneously with $1/k_{0}$ entering the sonic horizon $r_{s}:=\sqrt{w}/aH$. 
The scale factor is related to the horizon entry time by
    \begin{align}
        \frac{a(\eta_{\textmd{ta}})}{a(\eta_{\textmd{H}})}=\qty(\frac{\eta_{\textmd{ta}}}{\eta_{\textmd{H}}})^{\frac{2}{1+3w}}\simeq\qty(\frac{\eta_{k_{0}\textmd{S}}}{\sqrt{6}\eta_{k_{0}\textmd{H}}})^{\frac{2}{1+3w}}=\qty(\frac{1}{6w})^{\frac{1}{1+3w}}, \label{eq:aratio}
      \end{align} 
where the relations $a_{b}\propto\eta^{\frac{2}{1+3w}}, \eta_{\textmd{H}}=\sqrt{6}\eta_{k_{0}\textmd{H}}$ and $H^{-1}(\eta_{\textmd{H}})\simeq a(\eta_{\textmd{H}})r_{0}$ have been used.
Then, the $w$-dependence of the RMS of the spin can be explained as follows.
As $w$ decreases, the comoving radius of the sonic horizon decreases and the time of the sonic horizon entry is delayed. 
Then the ratio between the values of the scale factor at the turn-around and the horizon entry increases (Fig.~\ref{fig:aratio}), so that the RMS of the spin increases as $\sqrt{\langle a_{*}^2\rangle}\propto\qty(a(\eta_{\textmd{ta}})/a(\eta_{\textmd{H}}))^{1+2w}$.
The $w$-dependence of the sonic horizon entry would be overlooked if we simply assumed that turn-around occurs at almost the same time as the Hubble horizon entry of $r_{0}$, as in Ref.~\cite{DeLuca:2019buf}.

For the region $0.23\leq w\leq 1/3$, which corresponds to the QCD phase transition, we obtain $2.3\times10^{-3}\lesssim \sqrt{\langle a_{*}^2\rangle}\lesssim2.6\times10^{-3}$. 
Thus the QCD phase transition may not significantly amplify the value of the PBH spin.

It should be noted that we obtain $\sqrt{\langle a_{*}^2\rangle}>1$ for $0<w\lesssim 10^{-6}$ and $\sqrt{\langle a_{*}^2\rangle}\rightarrow\infty$ for $w\rightarrow0$. 
However, this result does not necessarily mean the possible generation of the over-spinning Kerr spacetime. 
Because, for $w\simeq0$, the pressure gradient, which we assumed as a main factor to impede PBH formation, should be negligible and a different treatment must be employed as in Ref.~\cite{Harada:2017fjm}. 
In addition, if the threshold of the amplitude for the PBH formation is sufficiently small, the high-peak assumption would be violated. 
Moreover, even if the overdense region has a supercritical Kerr parameter, the resultant black hole may be subcritical in the sense of the cosmic censorship.
Therefore, the result for $w\leq O(10^{-6})$ would not be reliable.

  \subsubsection{Mass-spin distributions with critical behavior}
\label{Respam}

Let us evaluate the distribution defined by Eq.~\eqref{eq:Prob} at the turn-around time $\eta_{\textmd{ta}}$ as follows. 
For $1<a_{*}$, which would be forbidden due to the effect of the stronger centrifugal force, we set $\nu_{\textmd{th}}=\infty$.
For $0\leq a_{*}\leq 1$, we fix $\nu_{\textmd{th}}=8$ and $\gamma=0.85$. 
The value of $\delta_{\textmd{pk}}(\eta_{H})$ is fixed by using Eq.~\eqref{eq:THC}. 

\begin{figure}[htbp]
    \begin{center}
     \includegraphics[clip,width=8cm]{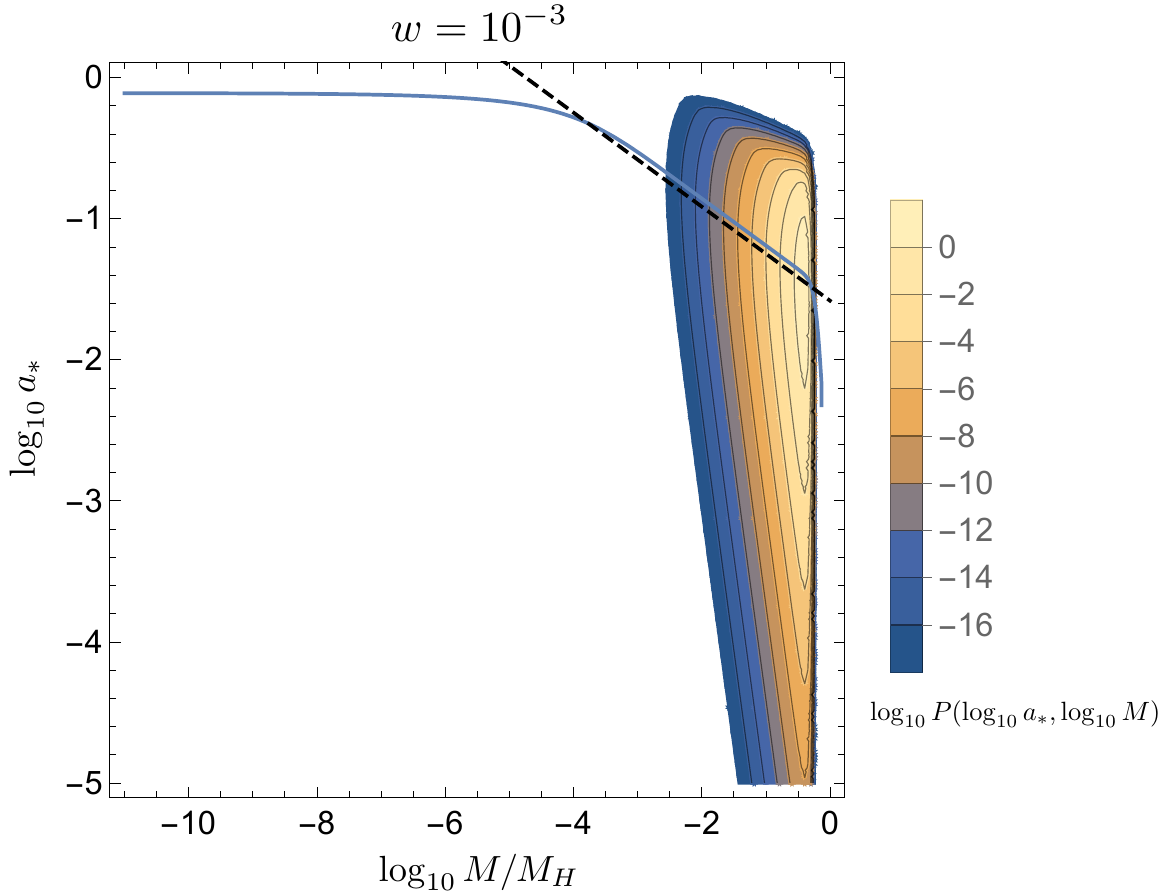}
     \includegraphics[clip,width=8cm]{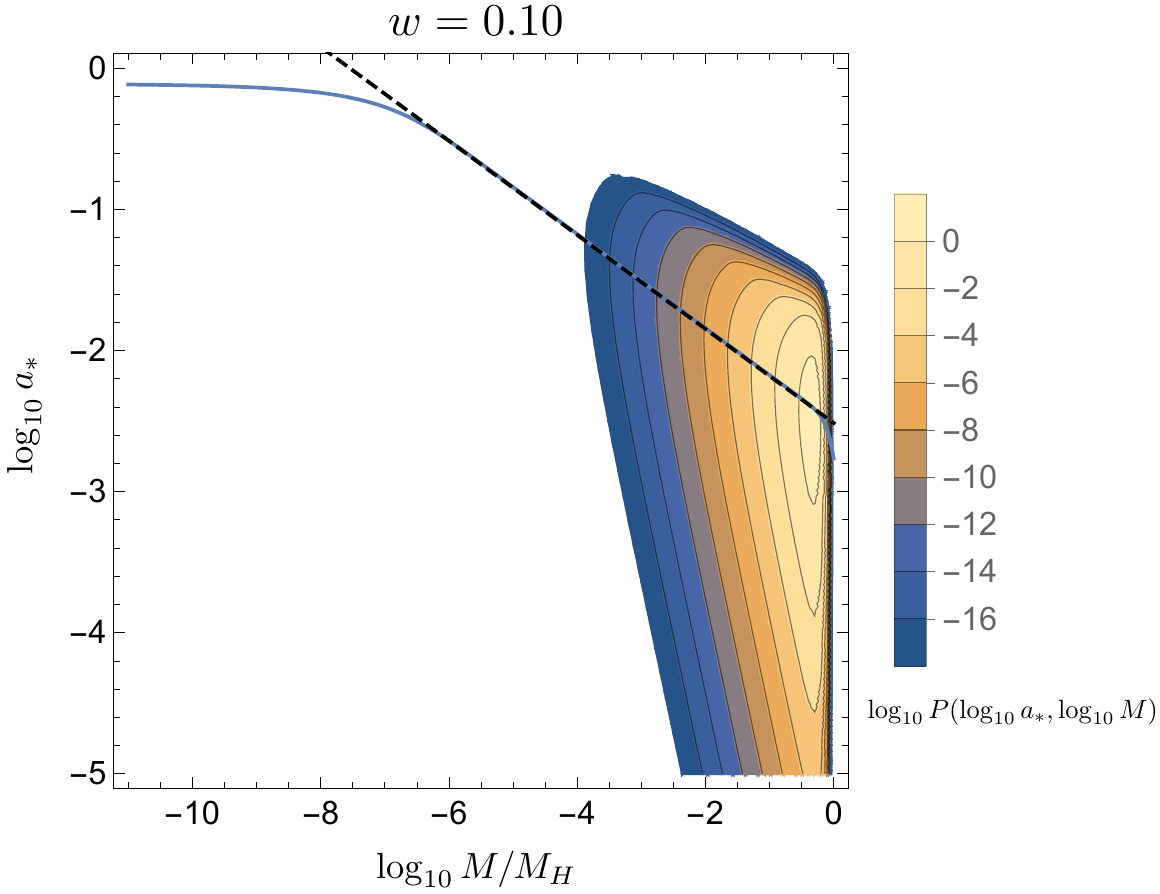}
     \includegraphics[clip,width=8cm]{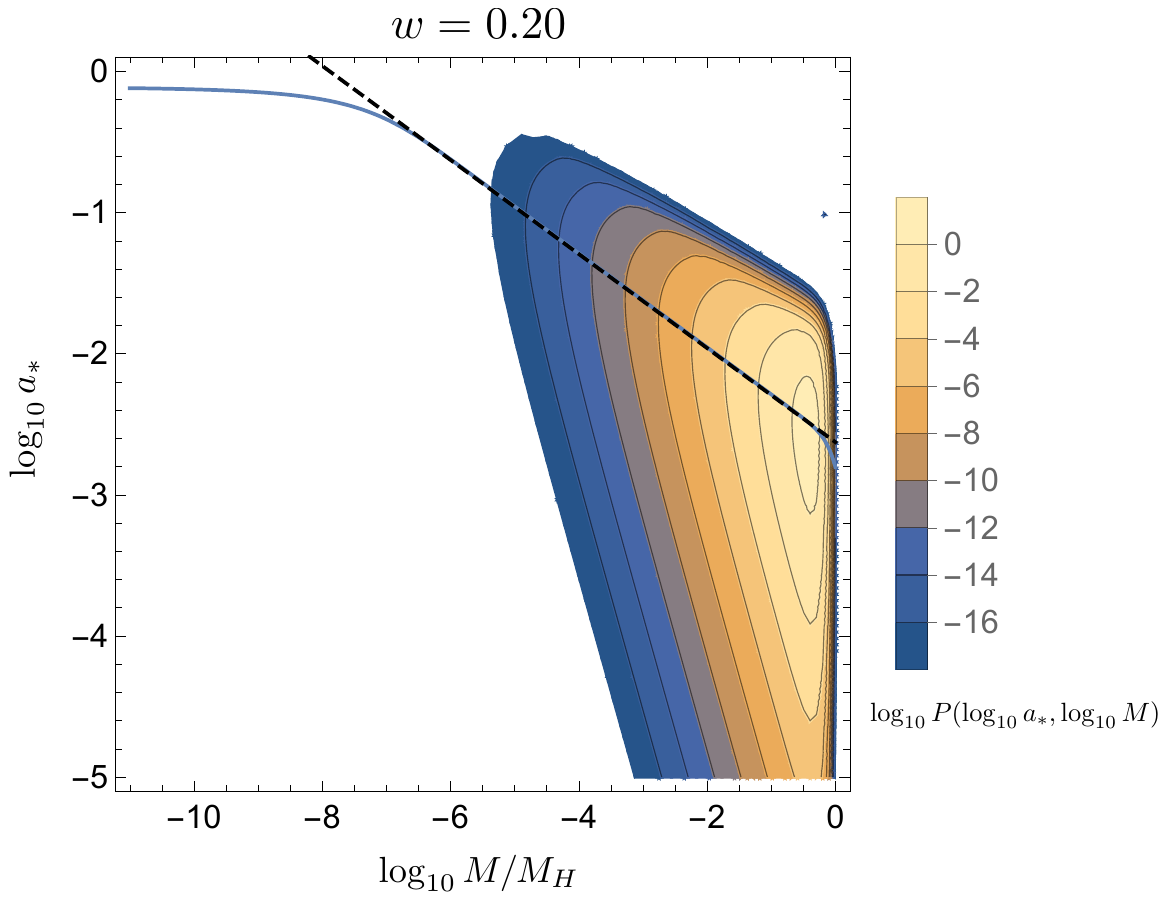}
     \includegraphics[clip,width=8cm]{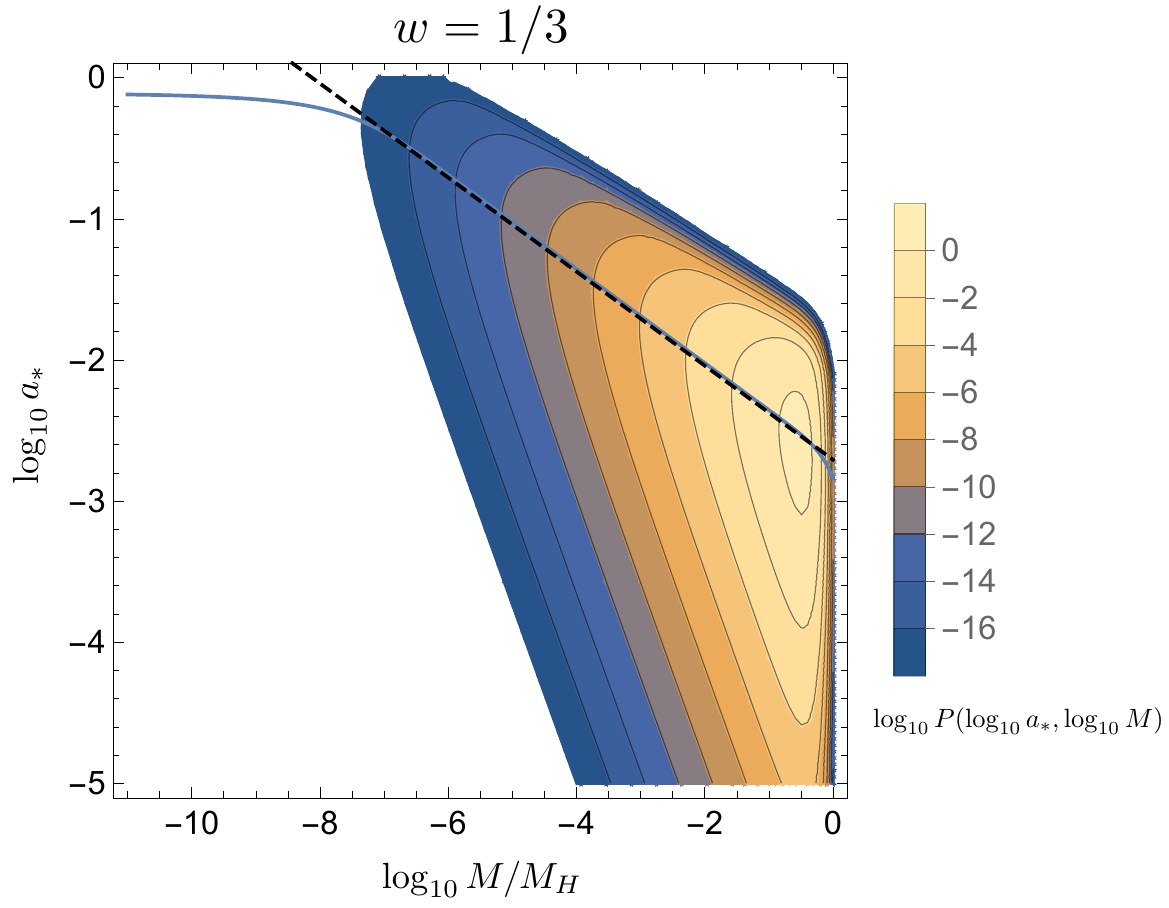}
    \caption{The contour plots of the mass-spin distributions $\log_{10}P(a_{*},M)$ for $w=10^{-3}$, $0.10$, $0.20$ and $1/3$.
    We have fixed as $\nu_{\textmd{th}}=8$ and $\gamma=0.85$ for $0\leq a_{*}\leq 1$ and $\nu_{\textmd{th}}=\infty$ for $1<a_{*}$.
    The black dot line shows the mean value of the Kerr parameter $\sqrt{\langle a^2_{*}\rangle}_{P}=\qty[\int_{0}^{1}da_{*} a^2_{*}P(a_{*}, M)/\int_{0}^{1}da_{*} P(a_{*}, M)]^{1/2}$ and the blue line shows its fitting $\sqrt{\langle a^2_{*}\rangle}_{P}\propto (M/M_{\textmd{H}})^{-1/3}$.
    }\label{fig:contour}
  \end{center}
   \end{figure}

The probability distributions $\log_{10}P(\log_{10}a_{*},\log_{10}M)$ with $w=10^{-3}, 0.10, 0.20, 1/3$ are depicted in Fig.~\ref{fig:contour}. 
The peak is around $10^{-0.7}\leq M/M_{\textmd{H}}\leq 10^{-0.3}$ and $10^{-3.0}\leq a_{*}\leq 10^{-0.7}$ for $w=10^{-3}$, while in the case of radiation fluid $w=1/3$, they are mostly distributed in $10^{-1.6}\leq M/M_{\textmd{H}}\leq 10^{-0.2}$ and $10^{-3.9}\leq a_{*}\leq 10^{-1.8}$.
We can see that, for smaller $w$, the distribution gets narrower for $M$ and a PBH is more likely to have a larger mass.
We note that this $w$-dependence of the distribution for $M$ is determined by the exponent $\kappa$ in the critical behavior~\eqref{eq:crit}.

In Fig.~\ref{fig:contour}, we also plot the mean value of $a_{*}$ defined by $\sqrt{\langle a^2_{*}\rangle}_{P}:=\qty[\int_{0}^{1}da_{*} a^2_{*}P(a_{*},M)/\int_{0}^{1}da_{*} P(a_{*},M)]^{1/2}$ (blue line) and its fitting $\sqrt{\langle a^2_{*}\rangle}_{P}\propto (M/M_{\textmd{H}})^{-1/3}$ (black dashed line). 
Note the difference between $\sqrt{\langle a^2_{*}\rangle}_{P}$ and $\sqrt{\langle a^2_{*}\rangle}$, which has been defined in Eq.~\eqref{eq:Sqa}.
Unlike $\sqrt{\langle a^2_{*}\rangle}$, $\sqrt{\langle a^2_{*}\rangle}_{P}$ is the mean value for the Kerr parameter in the range $0<a_{*}<1$, which can be interpreted as the PBH spin parameter.
We can see that, the mean value obeys the power-law as expected from Eq.~\eqref{eq:Sqa} for $a_{*}<10^{-0.5}$. 
However, the power-law is violated for $10^{-0.5}\leq a_{*}<1$. 
This is because we have assumed that the perturbation will not collapse for $1<a_{*}$.
We see that the value of $\sqrt{\langle a^2_{*}\rangle}_{P,M_{\textmd{H}}}$, defined by $\sqrt{\langle a^2_{*}\rangle}_{P}$ for $M\rightarrow M_{\textmd{H}}$, decreases with $w$. For example, from the fitting function, $\sqrt{\langle a^2_{*}\rangle}_{P,M_{\textmd{H}}}\simeq2.0\times10^{-3}$ for $w=1/3$ and $\sqrt{\langle a^2_{*}\rangle}_{P,M_{\textmd{H}}}\simeq2.6\times10^{-2}$ for $w=10^{-3}$.
These results are qualitatively consistent with the result in Sec.~\ref{Resrms}.
       
 \section{Summary and discussion}
 \label{Sum}

In this paper, we have evaluated the linear order effect on the spins of the PBHs formed in the universe dominated by a perfect fluid with $0< w\leq 1/3$. 
In the evaluation, we have assumed that the curvature perturbation obeys the Gaussian statistics and used the peak theory~\cite{Bardeen:1985tr, Heavens:1988}. 
We have also assumed that the density fluctuation has a high peak and applied the approximate probability distribution in the high-peak limit. 
We have focused on cases with a nearly monochromatic power spectrum.

We have found that the RMS of the spin $\sqrt{\langle a_{*}^2\rangle}$ decreases with $w$. 
One important factor which characterizes the $w$-dependence of $\sqrt{\langle a_{*}^2\rangle}$ is 
the ratio between the values of the scale factor at the horizon entry and the turn-around time. 
In our criterion of the turn-around time given in Eq.~\eqref{eq:TACond}, the turn-around time is very close to the time of the sonic horizon entry of the inverse wave number. 
Since the sonic horizon entry is delayed for a smaller value of $w$, the scale factor at the turn-around time increases, and then the value of $\sqrt{\langle a_{*}^2\rangle}$ increases. 
In the case of $M\sim M_{\textmd{H}}$, we have obtained $\sqrt{\langle a_{*}^2\rangle}=O(10^{-3})$ for the radiation-dominated universe and $\sqrt{\langle a_{*}^2\rangle}\geq O(10^{-2})$ for $w\leq0.01$.
We also found that the RMS is fitted by the power-law $\sqrt{\langle a_{*}^2\rangle}\propto w^{-0.49}$.
While $\sqrt{\langle a_{*}^2\rangle}$ drastically increases if $w$ is decreased from $w=1/3$ to $w\sim 0$, we have seen no significant change in the range corresponding
to the QCD crossover $(0.23\leq w\leq 1/3)$: $\sqrt{\langle a_{*}^2\rangle}\simeq2.3\times10^{-3}$ for $w=1/3$ and $\sqrt{\langle a_{*}^2\rangle}\simeq2.6\times10^{-3}$ for $w=0.23$.
This result indicates that the QCD phase transition does not significantly affect the value of the PBH spin.
We have also found that $\sqrt{\langle a_*^2\rangle}\propto(M/M_{\textmd{H}})^{-1/3}$, then the spin will be one order of magnitude larger for $M\sim 10^{-3}\times M_{\textmd{H}}$.

We have also discussed the $w$-dependence of the probability distribution $P(a_{*}, M)$ by considering the critical behavior~\cite{PhysRevD.59.104008}. 
We have found that, while the mass and spin are mostly distributed around $M\sim O(10^{-0.5})$ and $a_{*}\sim O(10^{-2.5})$ for $w=1/3$, the peak is around $M\sim O(10^{-0.4})$ and $a_{*}\sim O(10^{-1.5})$ for $w= 10^{-3}$.
 
We should admit that there is some uncertainty in the estimation.
We have assumed $\nu_{\textmd{th}}\gg1$ regardless of the value of $w$ and the collapsing region is nearly spherical.
However, for $w\simeq0$, PBHs could be formed from highly nonspherical perturbations~\cite{Khlopov:1980mg,Polnarev1982DustlikeSI,Harada:2016mhb}, and it should be important to study the effects of nonsphericity on tidal torque.
Also, the behavior of the fluctuation inside the Hubble horizon is highly nonlinear and we need a justification for determining the turn-around time based on the linear velocity perturbation.
The assumption of the conservation of the spin after the turn-around is also nontrivial.
They should be verified by numerical simulations.
Besides, it is reported in Refs.~\cite{Papanikolaou:2022cvo,Escriva:2022bwe,Musco:2023dak} that the time variation of the EoS parameter during the QCD phase transition could affect the threshold for PBH formation, and we should be careful when applying our results to cases with dynamical EoS parameters.
Although we have assumed a narrow power spectrum, estimation with other power spectra may be meaningful since the factor $\sqrt{1-\gamma^2}$ could make a larger contribution to the spin. 
In addition to that, throughout this paper, we have focused on the simplest formation scenario in which density fluctuations directly collapse to form PBHs against pressure gradient as the only factor that impedes the collapse.
Therefore, the applicability of the present results for other formation scenarios (e.g.~\cite{Deng:2017uwc,Kawana:2022lba}) is not clear.
Indeed, it is reported in Ref.~\cite{Flores:2021tmc} that the evolution of the PBH spins strongly depends on their cosmological origin.
Such issues are left for future work.

\section*{Acknowledgements}

The authors are very grateful to A. Escriv\'{a} for giving the results of numerical calculations.
The author (DS) would like to take this opportunity to thank the “Nagoya University Interdisciplinary Frontier Fellowship” supported by Nagoya University and JST, the establishment of university fellowships towards the creation of science technology innovation, Grant Number JPMJFS2120.
 This work was partially supported by JSPS KAKENHI Grant Numbers JP19H01895 (TH, CY, YK), JP19K03876(TH), JP20H05850 (CY), JP20H05853 (TH, CY, YK), JP21K20367
(YK).

 \appendix

 \section{Cosmological linear perturbations}
\label{Pert}

In this appendix, we show the solution of the linear perturbation which is necessary for the discussion, based on Ref.~\cite{Kodama:1984ziu}.

We assume that the background spacetime is given by a flat FLRW spacetime
\begin{align}
    &ds^2=-a^2d\eta^2+a^2\tilde{\gamma}_{ij}dx^{i}dx^{j}, \\
    &\tilde{\gamma}_{ij}dx^{i}dx^{j}=dr^2+r^2d\Omega^2,
\end{align}
and express scalar perturbations using the following harmonic functions:
\begin{align}
    Y,\quad Y_{i}=-\frac{1}{k}\nabla_{i}Y, \quad Y_{ij}=\frac{1}{k^2}\qty(\nabla_{j}\nabla_{i}Y-\frac{1}{3}\tilde{\gamma}_{ij}\Delta Y),
    \end{align}
where $Y$ is the scalar harmonic that satisfies
\begin{align}
    (\Delta+k^2)Y=0,
    \end{align}
and $\nabla_{i}$ denotes the covariant derivative with respect to $\tilde{\gamma}_{ij}$, and $\Delta:=\nabla_{i}\nabla^{i}$.

We define the metric perturbation using the components of Eq.~\eqref{eq:pert} as
\begin{align}
    &\alpha=a\qty(1+A_{\vec{k}}Y), \\
    &\beta_{i}=-a^2B_{\vec{k}}Y_{i}, \\
    &\gamma_{ij}=\tilde{\gamma}_{ij}+2H_{L,\vec{k}}Y\tilde{\gamma}_{ij}+2H_{T,\vec{k}}Y_{ij},
    \end{align}
and the perturbation of the extrinsic curvature $\mathcal{K}_{g,\vec{k}}$ as
\begin{align}
    K=K_{b}\qty(1+\mathcal{K}_{g,\vec{k}}Y),
\end{align}
where $K_{b}$ is the extrinsic curvature of the background.

We introduce the perturbation of the matter as
\begin{align}
    &T^{0}_{\ 0}=-\rho\qty[1+\delta_{\vec{k}}Y], \\
    &T^{0}_{\ j}=(\rho-p)\qty(v_{\vec{k}}-B_{\vec{k}})Y_{j}, \\
    &T^{i}_{\ i}=p\qty[\delta^{i}_{\ j}+\pi_{L,{\vec{k}}}\delta^{i}_{\ j}+\pi_{T,{\vec{k}}}Y^{i}_{\ j}].
\end{align}

Let us consider a gauge transformation defined by 
\begin{align}
    &\eta\rightarrow\bar{\eta}=\eta+T(\eta)Y, \\
    &x^{i}\rightarrow\bar{x}^{i}=x^{i}+L(\eta)Y^{i},
    \end{align}
where $T$ and $L$ are functions of $\eta$. Under this transformation, the perturbations transform as
\begin{align}
    &\bar{A}_{\vec{k}}=A_{\vec{k}}-T'-\frac{a'}{a}T, \quad \bar{B}_{\vec{k}}=B_{\vec{k}}+L'+kT, \quad \bar{H}_{L,\vec{k}}=H_{L,\vec{k}}-\frac{k}{3}L-\frac{a'}{a}T, \quad \bar{H}_{T,\vec{k}}=H_{T,\vec{k}}+kL, \\
    &\bar{\delta}_{\vec{k}}=\delta_{\vec{k}}+3(1+w)\frac{a'}{a}T, \quad \bar{v}_{\vec{k}}=v_{\vec{k}}+L', \quad \bar{\pi}_{L,\vec{k}}=\pi_{L,\vec{k}}+3\frac{c^2_{s}}{w}(1+w)\frac{a'}{a}T,
    \end{align}
where $c^2_{s}=\frac{p'}{\rho'}$ is the sound speed of the fluid and the dash denotes the derivative with respect to $\eta$.

We can obtain gauge invariant quantities by taking appropriate linear combinations of the perturbations. 
In this paper, we introduce the following two gauge invariants
\begin{align}
    \Delta_{\vec{k}}&:=\delta_{\vec{k}}+3(1+w)\frac{a'}{a}\frac{v_{\vec{k}}-B_{\vec{k}}}{k}, \\
    V_{\vec{k}}&:=v_{\vec{k}}-\frac{H'_{T,\vec{k}}}{k}.
    \end{align}

In the following, we assume that $w$ is a positive constant. 
We also assume that the perturbations are adiabatic and $\pi_{T}=0$. 
Then, we obtain the following equations for the gauge invariants:
\begin{align}
    &\Delta'_{\vec{k}}-3w\frac{a'}{a}\Delta_{\vec{k}}=-(1+w)kV_{\vec{k}}, \\
    &V'_{\vec{k}}+\frac{a'}{a}V_{\vec{k}}=-k\qty[\frac{4\pi\rho a^2}{k^2}-\frac{w}{1+w}]\Delta_{\vec{k}}.
    \label{eq:Euler}
    \end{align}
We can solve the equations as
\begin{align}
    \Delta_{\vec{k}}&=w^{(\beta-2)/2}z^{2-\beta}\qty[D(w)j_{\beta}(z)+E(w)n_{\beta}(z)], \\
    V_{\vec{k}}&=\frac{3}{2}\beta w^{(\beta-1)/2}z^{1-\beta}\qty[\qty{D(w)j_{\beta}(z)+E(w)n_{\beta}(z)}-\frac{z}{\beta+1}\qty{D(w)j_{\beta-1}(z)+E(w)n_{\beta-1}(z)}],
    \end{align}
where $z:=\sqrt{w}k\eta$, $\beta:=\frac{2}{3w+1}$, $D(w)$ and $E(w)$ are constants of integration. 
$j_{\beta}(z)$ and $n_{\beta}(z)$ are the spherical Bessel functions and in the long-wavelength regime $z\ll1$ and $j_{\beta}$ correspond to the growing solution and the decaying solution, respectively. 
In the following, we focus on the growing mode and approximate the solutions as
\begin{align}
    \Delta_{\vec{k}}&\simeq D(w)w^{(\beta-2)/2}z^{2-\beta}j_{\beta}(z), \\
    V_{\vec{k}}&\simeq\frac{3}{2}D(w)\beta w^{(\beta-1)/2}z^{1-\beta}\qty[j_{\beta}(z)-\frac{z}{\beta+1}j_{\beta-1}(z)].
\end{align}

In the CMC (uniform Hubble) slicing $\qty(\mathcal{K}_{g,\vec{k}}=0)$ with a gauge such that $B_{\vec{k}}=0$, the density perturbation, the velocity and the intrinsic curvature are written as
\begin{align}
    \delta_{\vec{k},\textmd{CMC}}
    &=\Delta_{\vec{k}}-\frac{6\beta\sqrt{w}(1+w)z}{2z^2+9\beta^2w(1+w)}V_{\vec{k}}, \label{eq:delCMC}  \\
    v_{\vec{k},\textmd{CMC}}
    &=\frac{2z^2}{2z^2+9\beta^2w(1+w)}V_{\vec{k}}, \\
    \mathcal{R}_{\vec{k},\textmd{CMC}}
    &=\frac{3\beta^2w}{2z^2}\qty(\Delta_{\vec{k}}-\frac{6\beta\sqrt{w}(1+w)z}{2z^2+9\beta^2w(1+w)}V_{\vec{k}}),
    \end{align}
where 
\begin{align}
    \mathcal{R}_{\vec{k}}:=H_{L,\vec{k}}+\frac{H_{T,\vec{k}}}{3}
    \end{align}
is the perturbation of the intrinsic curvature.

In the conformal Newtonian gauge $(B_{\vec{k}}=H_{T,\vec{k}}=0)$, we obtain the density perturbation and the velocity as
\begin{align}
    \delta_{\vec{k},\textmd{CN}}
    &=\Delta_{\vec{k}}-\frac{3\sqrt{w}(1+w)\beta}{z}V_{\vec{k}},  \\
    v_{\vec{k},\textmd{CN}}&=V_{\vec{k}}.
    \label{eq:vCN}
    \end{align}

We can define another gauge invariant
\begin{align}
    \Phi_{\vec{k}}:=-H_{L,\vec{k}}-\frac{H_{T,\vec{k}}}{3}-\frac{1}{k}\frac{a'}{a}\qty(B_{\vec{k}}-\frac{H'_{T,\vec{k}}}{k}),
    \end{align}
which is related to $\Delta_{\vec{k}}$ by
\begin{align}
    \Phi_{\vec{k}}&=-\frac{3\beta^2w}{2z^2}\Delta_{\vec{k}},
    \end{align}
and define the transfer functions by
\begin{align}
    &v_{\vec{k},\textmd{CMC}}=T_{v_{\textmd{CMC}}}(\vec{x},\eta) \Phi_{\vec{k}}(0), \\
    &v_{\vec{k},\textmd{CN}}=T_{v_{\textmd{CN}}}(\vec{x},\eta) \Phi_{\vec{k}}(0).
    \end{align}

\bibliography{bibs/hoge}

\begin{thebibliography}{10}

\bibitem{Carr:2017jsz}
Bernard Carr, Martti Raidal, Tommi Tenkanen, Ville Vaskonen, and Hardi
  Veerm\"ae.
\newblock {Primordial black hole constraints for extended mass functions}.
\newblock {\em Phys. Rev. D}, 96(2):023514, 2017.

\bibitem{Carr:2020gox}
Bernard Carr, Kazunori Kohri, Yuuiti Sendouda, and Jun'ichi Yokoyama.
\newblock {Constraints on primordial black holes}.
\newblock {\em Rept. Prog. Phys.}, 84(11):116902, 2021.

\bibitem{LIGOScientific:2020stg}
R.~Abbott et~al.
\newblock {GW190412: Observation of a Binary-Black-Hole Coalescence with
  Asymmetric Masses}.
\newblock {\em Phys. Rev. D}, 102(4):043015, 2020.

\bibitem{Carr:1974nx}
Bernard~J. Carr and S.~W. Hawking.
\newblock {Black holes in the early Universe}.
\newblock {\em Mon. Not. Roy. Astron. Soc.}, 168:399--415, 1974.

\bibitem{Harada:2013epa}
Tomohiro Harada, Chul-Moon Yoo, and Kazunori Kohri.
\newblock {Threshold of primordial black hole formation}.
\newblock {\em Phys. Rev. D}, 88(8):084051, 2013.
\newblock [Erratum: Phys.Rev.D 89, 029903 (2014)].

\bibitem{Escriva:2019phb}
Albert Escriv\`a, Cristiano Germani, and Ravi~K. Sheth.
\newblock {Universal threshold for primordial black hole formation}.
\newblock {\em Phys. Rev. D}, 101(4):044022, 2020.

\bibitem{Escriva:2020tak}
Albert Escriv\`a, Cristiano Germani, and Ravi~K. Sheth.
\newblock {Analytical thresholds for black hole formation in general
  cosmological backgrounds}.
\newblock {\em JCAP}, 01:030, 2021.

\bibitem{Shibata:1999zs}
Masaru Shibata and Misao Sasaki.
\newblock {Black hole formation in the Friedmann universe: Formulation and
  computation in numerical relativity}.
\newblock {\em Phys. Rev. D}, 60:084002, 1999.

\bibitem{Musco:2004ak}
Ilia Musco, John~C. Miller, and Luciano Rezzolla.
\newblock {Computations of primordial black hole formation}.
\newblock {\em Class. Quant. Grav.}, 22:1405--1424, 2005.

\bibitem{Harada:2015yda}
Tomohiro Harada, Chul-Moon Yoo, Tomohiro Nakama, and Yasutaka Koga.
\newblock {Cosmological long-wavelength solutions and primordial black hole
  formation}.
\newblock {\em Phys. Rev. D}, 91(8):084057, 2015.

\bibitem{Chiba:2017rvs}
Takeshi Chiba and Shuichiro Yokoyama.
\newblock {Spin Distribution of Primordial Black Holes}.
\newblock {\em PTEP}, 2017(8):083E01, 2017.

\bibitem{DeLuca:2019buf}
V.~De~Luca, V.~Desjacques, G.~Franciolini, A.~Malhotra, and A.~Riotto.
\newblock {The initial spin probability distribution of primordial black
  holes}.
\newblock {\em JCAP}, 05:018, 2019.

\bibitem{Mirbabayi:2019uph}
Mehrdad Mirbabayi, Andrei Gruzinov, and Jorge Nore\~na.
\newblock {Spin of Primordial Black Holes}.
\newblock {\em JCAP}, 03:017, 2020.

\bibitem{He:2019cdb}
Minxi He and Teruaki Suyama.
\newblock {Formation threshold of rotating primordial black holes}.
\newblock {\em Phys. Rev. D}, 100(6):063520, 2019.

\bibitem{Harada:2020pzb}
Tomohiro Harada, Chul-Moon Yoo, Kazunori Kohri, Yasutaka Koga, and Takeru
  Monobe.
\newblock {Spins of primordial black holes formed in the radiation-dominated
  phase of the universe: first-order effect}.
\newblock {\em Astrophys. J.}, 908(2):140, 2021.

\bibitem{Chongchitnan:2021ehn}
Siri Chongchitnan and Joseph Silk.
\newblock {Extreme-value statistics of the spin of primordial black holes}.
\newblock {\em Phys. Rev. D}, 104(8):083018, 2021.

\bibitem{Harada:2017fjm}
Tomohiro Harada, Chul-Moon Yoo, Kazunori Kohri, and Ken-Ichi Nakao.
\newblock {Spins of primordial black holes formed in the matter-dominated phase
  of the Universe}.
\newblock {\em Phys. Rev. D}, 96(8):083517, 2017.
\newblock [Erratum: Phys.Rev.D 99, 069904 (2019)].

\bibitem{Borsanyi:2016ksw}
Sz. Borsanyi et~al.
\newblock {Calculation of the axion mass based on high-temperature lattice
  quantum chromodynamics}.
\newblock {\em Nature}, 539(7627):69--71, 2016.

\bibitem{Bardeen:1985tr}
James~M. Bardeen, J.~R. Bond, Nick Kaiser, and A.~S. Szalay.
\newblock {The Statistics of Peaks of Gaussian Random Fields}.
\newblock {\em Astrophys. J.}, 304:15--61, 1986.

\bibitem{Heavens:1988}
A.~Heavens and J.~Peacock.
\newblock {Tidal torques and local density maxima}.
\newblock {\em Mon.Not.Roy.Astron.Soc.}, 232:339, 1988.

\bibitem{Yoo:2018kvb}
Chul-Moon Yoo, Tomohiro Harada, Jaume Garriga, and Kazunori Kohri.
\newblock {Primordial black hole abundance from random Gaussian curvature
  perturbations and a local density threshold}.
\newblock {\em PTEP}, 2018(12):123E01, 2018.

\bibitem{Koga:2022bij}
Yasutaka Koga, Tomohiro Harada, Yuichiro Tada, Shuichiro Yokoyama, and
  Chul-Moon Yoo.
\newblock {Effective Inspiral Spin Distribution of Primordial Black Hole
  Binaries}.
\newblock {\em Astrophys. J.}, 939(2):65, 2022.

\bibitem{PhysRevD.59.104008}
Tatsuhiko Koike, Takashi Hara, and Satoshi Adachi.
\newblock Critical behavior in gravitational collapse of a perfect fluid.
\newblock {\em Phys. Rev. D}, 59:104008, Apr 1999.

\bibitem{Albert}
{Private communication with Albert Escriv\'{a}}.

\bibitem{Khlopov:1980mg}
M.~Yu. Khlopov and A.~G. Polnarev.
\newblock {PRIMORDIAL BLACK HOLES AS A COSMOLOGICAL TEST OF GRAND UNIFICATION}.
\newblock {\em Phys. Lett. B}, 97:383--387, 1980.

\bibitem{Polnarev1982DustlikeSI}
Alexander~G. Polnarev and Maxim~Yu. Khlopov.
\newblock Dustlike stages in the early universe, and constraints on the
  primordial black hole spectrum.
\newblock 1982.

\bibitem{Harada:2016mhb}
Tomohiro Harada, Chul-Moon Yoo, Kazunori Kohri, Ken-ichi Nakao, and Sanjay
  Jhingan.
\newblock {Primordial black hole formation in the matter-dominated phase of the
  Universe}.
\newblock {\em Astrophys. J.}, 833(1):61, 2016.

\bibitem{Papanikolaou:2022cvo}
Theodoros Papanikolaou.
\newblock {Toward the primordial black hole formation threshold in a
  time-dependent equation-of-state background}.
\newblock {\em Phys. Rev. D}, 105(12):124055, 2022.

\bibitem{Escriva:2022bwe}
Albert Escriv\`a, Eleni Bagui, and Sebastien Clesse.
\newblock {Simulations of PBH formation at the QCD epoch and comparison with
  the GWTC-3 catalog}.
\newblock {\em JCAP}, 05:004, 2023.

\bibitem{Musco:2023dak}
Ilia Musco, Karsten Jedamzik, and Sam Young.
\newblock {Primordial black hole formation during the QCD phase transition:
  threshold, mass distribution and abundance}.
\newblock 3 2023.

\bibitem{Deng:2017uwc}
Heling Deng and Alexander Vilenkin.
\newblock {Primordial black hole formation by vacuum bubbles}.
\newblock {\em JCAP}, 12:044, 2017.

\bibitem{Kawana:2022lba}
Kiyoharu Kawana, Philip Lu, and Ke-Pan Xie.
\newblock {First-order phase transition and fate of false vacuum remnants}.
\newblock {\em JCAP}, 10:030, 2022.

\bibitem{Flores:2021tmc}
Marcos~M. Flores and Alexander Kusenko.
\newblock {Spins of primordial black holes formed in different cosmological
  scenarios}.
\newblock {\em Phys. Rev. D}, 104(6):063008, 2021.

\bibitem{Kodama:1984ziu}
Hideo Kodama and Misao Sasaki.
\newblock {Cosmological Perturbation Theory}.
\newblock {\em Prog. Theor. Phys. Suppl.}, 78:1--166, 1984.

\end{thebibliography}
\bibliographystyle{unsrt.bst}

\end{document}